\documentclass[12pt]{article}

\usepackage{graphicx}

\begin{document}

\begin{center}

{\large\bf The anisotropy of two dimensional percolation clusters
of self-affine models}

\bigskip

Fatemeh Ebrahimi\\
{\it Department of Physics, University of Birjand, Birjand,
 Iran, 97175-615\\}

\end{center}

\begin{abstract}
The anisotropy parameter of two-dimensional equilibrium clusters
of site percolation process in long-range self-affine correlated
structures are studied numerically. We use a fractional Brownian
Motion(FBM) statistic to produce both persistent and
anti-persistent long-range correlations in 2-D models. It is seen
that self affinity makes the shape of percolation clusters
slightly more isotropic. Moreover, we find that the sign of
correction to scaling term is determined by the nature of
correlation. For persistent correlation the correction to scaling
term adds a negative contribution to the anisotropy of
percolation clusters, while for the anti-persistent case it is
positive.

\end{abstract}

\section{introduction}

The shape of clusters in percolation type models is an interesting
physical property. Numerical simulations on random models has
revealed that percolation clusters at the percolation threshold
are asymmetric~\cite{fam,s1,s2,Q,fe} which is a significant
result, especially because of the close connection between
percolation model and the theory of second-order phase
transition~\cite{fam,fis}.

Generally speaking, the shape of a 2-dimensional cluster is
determined by $R^2_{min}$ and $R^2_{max}$ ($R^2_{min} \leq
R^2_{max}$), the eigenvalues (the principal radii of gyration) of
the cluster radius of gyration tensor. If $R^2_{min}=R^2_{max}$,
the cluster is spherically symmetric. Otherwise, it is
anisotropic and we can probe the degree of its anisotropy by
defining a proper anisotropy quantifier. Family et al~\cite{fam},
proposed an convenient asymmetry measure,
$A_s=R^2_{min}/R^2_{max}$, called the anisotropy parameter of a
$s$-site cluster. The quantity $A_s$  when properly averaged over
all clusters with the same size is denoted by $\langle
A_s\rangle$ and is an estimate of the anisotropy parameter of
$s$-site clusters in the ensemble. The case $\langle
A_s\rangle=1$, corresponds to spherical symmetry. For anisotropic
objects, $\langle A_s\rangle$ is less than unity (the term
anisotropy parameter may be misleading; the shape of the cluster
is more isotropic for larger value of $A_s$). The asymptotic
behaviour of $\langle A_\infty \rangle$ is obtained by taking the
limit $s\rightarrow \infty$. Using this method for two
dimensional random percolation, they observed for the first time
that percolation clusters are not isotropic and estimated
$\langle A_\infty\rangle\cong0.4$ as the asymptotic value for the
anisotropy of infinitely large percolation clusters.

The most studied percolation systems have been involved with
totally random and un-correlated models.  But, in some important
practical applications of percolation theory, the nature of
disorder is not completely random and there are correlations in
the properties of the medium. For example, percolation theory has
been used to understanding and explaining some important aspects
of multi-phase flow in porous media~\cite{sa0}. However, studies
have demonstrated the existence of long-rang correlations in the
permeability distributions and porosity logs of some field scale
natural porous media like sedimentary rocks~\cite{h1,h2,sam}.

A common way of introducing long-range correlations into a medium
property is fractional Brownian motion (FBM) $\textbf{
D}_B(\overrightarrow{x})$~\cite{fbm1,fbm2}. FBM is an ergodic,
non-stationary stochastic process whose increments are
statistically self-similar such that its mean square fluctuation
is proportional to an arbitrary power of the spatial displacement
$\overrightarrow{x}$

\begin{equation}
\langle
\textbf{D}_B(\overrightarrow{x})-\textbf{D}_B(\textbf{0})]^2\rangle
\sim |\overrightarrow{x}|^{2H}
\end{equation}
$H$ is called the Hurst exponent and determines the type of
correlations. If $H=0.5$, the above equation produces the
ordinary Brownian motion, which means that in this case there is
no correlation between different increments. If $H>0.5$, then FBM
generates positive correlations, i.e. all the points in a
neighborhood of a given point obey more or less the same trend. If
$H<0.5$, FBM is anti-persistence, i.e. a trend at a point will
not be likely followed in its immediate neighborhood.

In this paper we address the shape of 2-D percolation clusters of
self-affine models by evaluating  their anisotropy parameters.The
reason that we have chosen FBM process is tenfold. First, FBM
generates long-range and at the same time isotropic correlations
in the field. Therefore, the host lattice retains its isotropy.
Second it has been demonstrated that such process has practical
applications in earth sciences and also reservoir engineering,
where the permeability field and also the porosity distribution
of many real oil reservoirs and aquifer follow FBM
statistic~\cite{mrs,fbm3}.  A FIB has been used by Schmittbuhl et
al~\cite{svr} and  Sahimi~\cite{sam} for generating a percolation
model with long-range correlations.

The paper is organized as follow. After describing the simulation
method in section 2, we present and discuss the results of our
extensive numerical simulations of percolation processes for both
random and self affine models in section 3, with concluding
remarks at section 4.

\section{ Simulation method}

We start with a $L\times L$ square lattice, and assign to each
lattice site $\overrightarrow{x_i}$  a random number drawn from a
normalized FBM distribution. There are a number of methods which
are capable of producing the FBM statistics~\cite{mrs,fbm2}. We
have used one of the most popular one, the method of fast Fourier
transformation (FFT) filtering which is based on the fact that the
power spectrum of FBM is given by:
\begin{equation}
\mathcal{S}(\mathbf{\omega})=\frac{a_0}{(\omega_x^2+\omega_y^2)^p}\
\end{equation}
where $a_0$ is a numerical constant,
$\mathbf{\omega}=(\omega_1,\omega_2)$, with $\omega_i$ being the
Forger component in the $i$th direction and $p=H+1$. In the FAT
method, one starts with a white noise $W(x,y)$ defined on the
lattice sites. The power spectrum of $W(x,y)$ is constant and
independent of frequency. Therefore, filtering  $W(x,y)$ with a
transfer function  $\sqrt{\mathcal{S}(\mathbf{\omega})}$ generates
another noise whose spectral density is proportional to
$\mathcal{S}(\mathbf{\omega})$. The method is straightforward and
fast, but it usually produces periodic noises. Therefore, one has
to produce a larger lattice and keep only a portion (typically
1/4 in two-dimensional lattices). In fig.1 two different
realizations of FIB on a $256\times 256$ square lattice have been
shown.

To generate percolation clusters one chooses a threshold $r$. All
lattice sites whose $\textbf{D}_B$ values are less than $r$ are
considered filled. The concentration $p$ of filled sites depends
on both  $\textbf{D}_B(\overrightarrow{x})$ and $r$ through:
\begin{equation}
p=\int_{0}^{r} \textbf{D}_B(\overrightarrow{x})d\overrightarrow{x}
\end{equation}
\begin{figure}

\begin{center}
\includegraphics[width= 10.8 truecm,height= 7.8truecm]{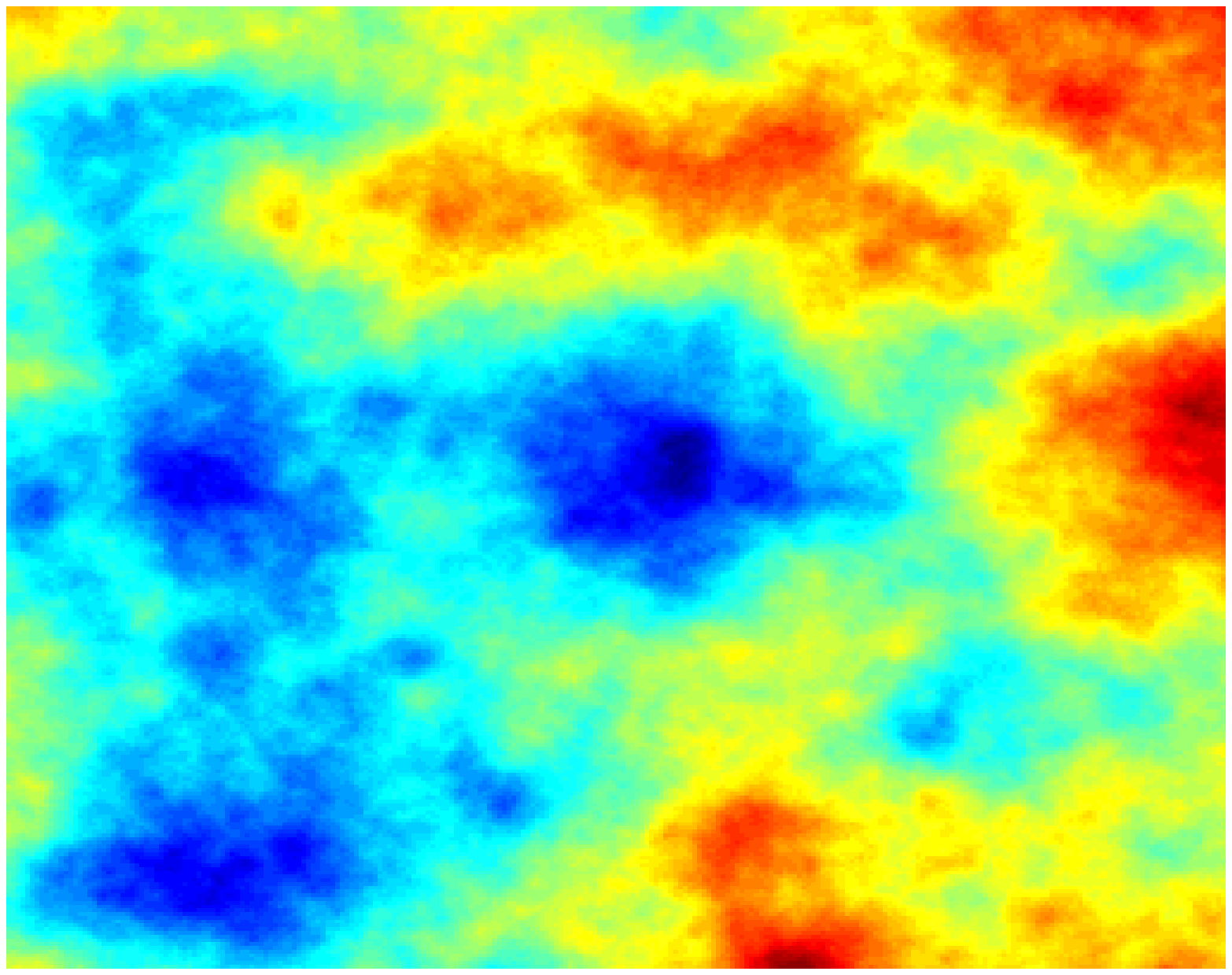}

\bigskip

%\end{center}
%\caption{a} %\label{fig.1a}
%\end {figure}
%\begin{figure}
%\begin{center}
\includegraphics[width= 11 truecm]{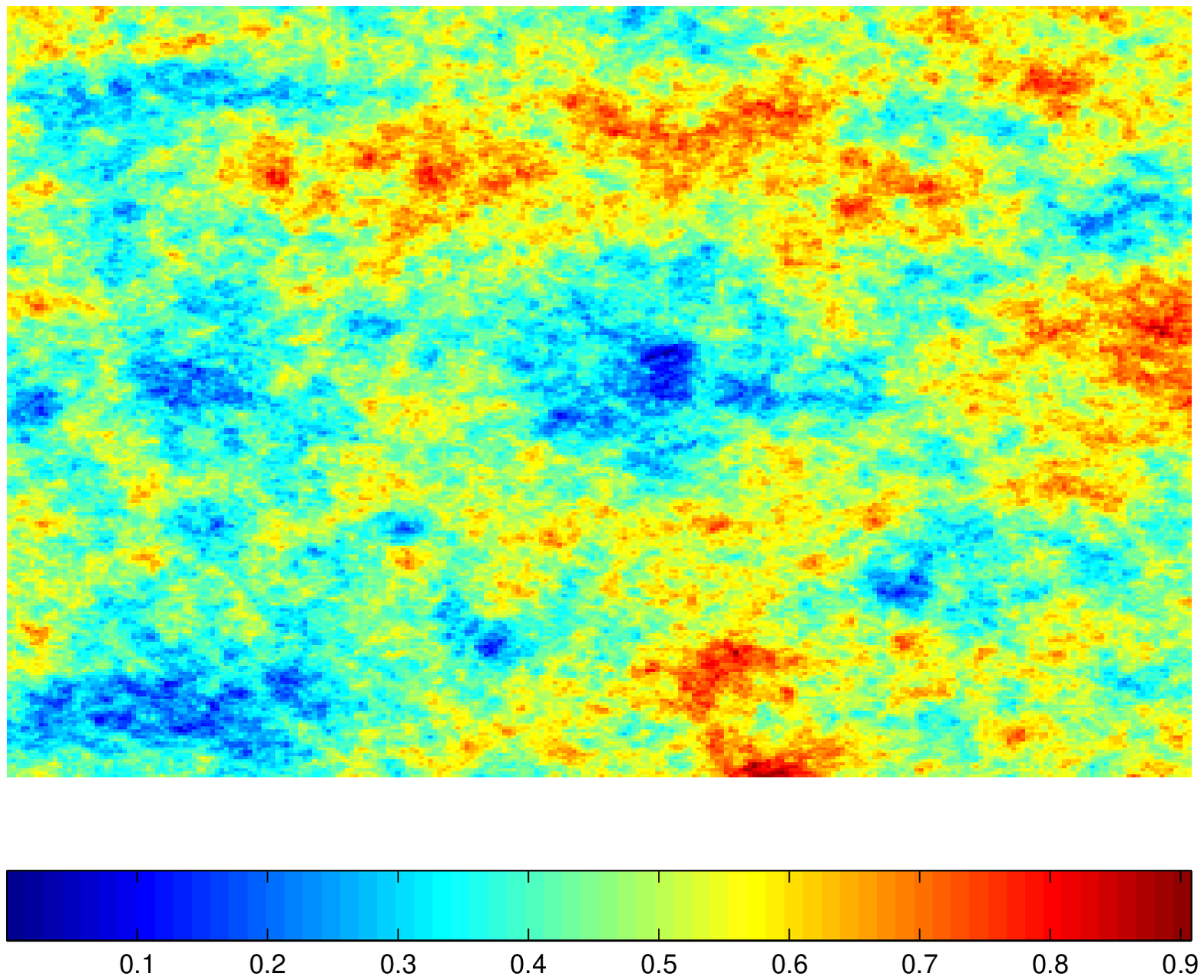}
\end{center}
\caption{\small{Two different realizations of FBM statistics on a
$256\times 256$ square lattice: $H=0.8$ (top) and $H=0.2$ (bottom)
}} \label{fig.1}
\end {figure}
The percolation threshold $p_c$ has been defined as the minimum
concentration $p$ of filled sites to form an infinite cluster
with probability $1$. It has been observed that percolation
transitions on long-range self-affine structures behaves
differently from ordinary percolation in some important
aspects~\cite{svr}. For example, Marrink et al~\cite{mpk} found
that percolation thresholds $p_c$ of self-affine lattices are
strongly dependent on the spanning rule employed even when the
lattice is effectively infinite. Following the work of Reynolds
et al~\cite{rea} on un-correlated square lattices, they
considered three different spanning rules  for percolation on
self-affine models, $ \mathcal R_0 $ , the probability of spanning
either horizontally or vertically or both, $ \mathcal R_1$, the
probability of spanning in a specified direction (e.g.
horizontally), and $ \mathcal R_2 $, the probability of spanning
both horizontally and vertically. The numerical results obtained
by Marrink et al established that unlike the urn-correlated random
model three critical concentrations: $p_c^{eith}$, $p_c^{spec}$,
and $p_c^{both}$ (corresponding to percolation rules  $ \mathcal
R_0 $,  $ \mathcal R_1 $,  and $ \mathcal R_2 $ respectively) do
not converge for infinite self-affine models. Therefore, the
definition of the percolation threshold depends on the desired
applications and one must consider the appropriate percolation
threshold to obtain critical parameters of percolation transition
of self affine models.

At each specified percolation threshold, there exists a
distribution of distinct, finite clusters (equilibrium
percolation clusters). In this work we made the search for
clusters  by using the Hoshen-Kopelman algorithm~\cite{hoko}. To
remove the effect of boundary conditions from our numerical
estimations, we only count those clusters which do not touch
lattice boundaries. We call them internal clusters. For each
internal cluster of an arbitrary size $s$, we evaluate  the
cluster radius of gyration tensor $\textbf{G}$
\begin{equation}
\textbf{G}=\sum_{i=1}^N
(\overrightarrow{x}_i^2\textbf{I}-\overrightarrow{x}_i
\overrightarrow{x}_i)
\end{equation}
In the above definition, $ \overrightarrow{x} _i$ is the distance
of occupied site $i$ from the cluster's center of mass and $s$ is
the size of the cluster.  The principal radii of gyration of the
cluster, $R_{min}^2$ and $R_{max}^2$, are obtained via
diagonalization of $\textbf{G}$. At the percolation threshold the
variations in the $R^2_i$, have the following asymptotic form:
\begin{equation}
\langle R^2_i \rangle= r_i
s^{2\nu}(1+a_is^{-\theta}+b_is^{-1}+...) \;\;\;\;\;\;\;\;\;\;\;\;
i=min\; or \;max
\end{equation}
where $\nu$ is the leading scaling exponent and $\theta$ is the
non-analytical correction-to-scaling exponent of the cluster. The
coefficients $r_i$, $a_i$, and $b_i$ are all independent of
$s$~\cite{fam}.  The shape of the cluster is then characterized
by evaluating its  anisotropy parameter, as described
previously.  Finally, we obtain the mean values of anisotropy
quantifier by averaging it over all the percolation clusters with
the same size $s$, resulted in different lattice realizations.
Then, the results have been lumped together at the centers of
blocks of size $[2^{m-1},2^m]$. This procedure not only helps to
eliminate correction-to-scaling for small clusters~\cite{Q}, but
it produces new data points which are usually less correlated
than the original data~\cite{fp}.

The most time-consuming step in these kind of simulations is
generation of FBM statistics itself. Generating FBM on large
lattices requires significant CPU time and computer memory. On
the other hand, in order to eliminate finite-size effects from
numerical results, we need lattices with large $L$ . In this work,
we have fixed the linear size of our square lattices to $L=256$
which is both computationally tractable and at the same time
large enough to make sure that our results are not affected by
finite-size effects~\cite{sam}. To achieve highly accurate
results, we estimate the mean values by sampling a large number
of lattice realizations ($30\; 000$ samples in each case).

In order to make an appropriate and reliable sampling of cluster
shapes in configurations space it is necessary that the medium
linear size be large enough such that all possible configurations,
including the most anisotropic ones potentially can happen (the
condition of effectively infinite large medium). This puts a
maximum on cluster size $s$. Indeed, for large clusters , i.e.
when $s>>L^2$, the condition of effectively infinite large medium
would not be fulfilled and as such, the sampling would be in
favor of more isotropic configurations.

\section{ Results}
Let us first look at the shape of percolation clusters in
completely random media. As mentioned in the first section, the
estimated value of $A_\infty$  at percolation threshold has been
reported by several authors. Here, we provide our estimations of
the anisotropy parameters of percolation clusters for some
selected values of $p$ in Table1. The total number of lattice
realizations has been $100\; 000$ in this case. The cluster
numbers is a function of $p$, but it always decreases rapidly with
$s$~\cite{st}. Therefore, when $|p-p_c|$ is not small, the total
number of clusters of larger sizes is not large enough to give a
reliable statistics and we do not include the anisotropy
parameters of such clusters in our reports.

%\begin{center}
\begin{table}
%\caption{\label{table1}The anisotropy parameters of random
%lattices at different concentration $p$ }%\\ %\hline
%\begin{indented}
%\bigskip
%\begin{ruledtabular}
% \caption{
{\small {\bf Table 1}.The anisotropy parameters of random model}
%\bigskip

\begin{tabular}{lllllllll}
\hline
$m$ & $$ & $p= 0.454$ & $$ & $ p=0.550$ & $$ & $p=0.593$ &
$$ & $p=0.650$
\\  \hline
\\
$3$ & $$ & $.340(4) $ & $ $& $.355(4) $ & $$ & $.362(5) $ & $$ & $.375(6)$  \\
$4$ & $$ & $.3307(2)$ & $ $& $.3467(2)$ & $$ & $.3545(3)$ & $$ & $.3685(5)$  \\
$5$ & $$ & $.3319(2)$ & $ $& $.3511(4)$ & $$ & $.3607(6)$ & $$ & $.3786(9)$  \\
$6$ & $$ & $.3330(1)$ & $ $& $.3565(3)$ & $$ & $.3688(4)$ & $$ & $.3914(7)$  \\
$7$ & $$ & $.3313(2)$ & $ $& $.3594(1)$ & $$ & $.3752(3)$ & $$ & $.405(1)$  \\
$8$ & $$ & $.326(1) $ & $ $& $.3591(1)$ & $$ & $.3794(2)$ & $$ & $.419(3)$  \\
$9$ & $$ & $---   $ & $ $& $.3569(2)$ & $$ & $.3813(3)$ & $$ & $---  $  \\
 \\ \hline
 \end{tabular}
%\end{ruledtabular}
%\end{indented}
\end{table}
%\end{center}

As seen from this results, in ordinary percolation the cluster
mean anisotropy parameters is a function of both $p$ and $s$.
This is natural as the number of a specific cluster configuration
in random percolation model is proportional to $p^s(1-p)^t$, where
$s$ is the cluster size and $t$ is its perimeter~\cite{st}. In
fact more careful investigations show some important trends in
the behaviour of $\langle A_s(p) \rangle$ as a function of $s$
and $p$. For example, looking at the table we can see that all
the values of $s$, the mean anisotropy parameter grows with $p$.
Meanwhile, the variation of $\langle A_s \rangle$ for a given $p$
is much more complicated. It seems that for $p<p_c$, $\langle
A_s\rangle$ has a maximum at some moderate value of $s$, while
when $p>p_c$, the larger clusters are always more symmetric than
smaller ones. The behaviour of $\langle A_s\rangle$ at
percolation threshold $p\simeq0.593$ is consistent with
equation2, i.e., it evolves to an asymptotic value at large
values of $s$~\cite{fam,s1,s2,Q,fe}.

\begin{figure}
\begin{center}
\includegraphics[width= 10 truecm]{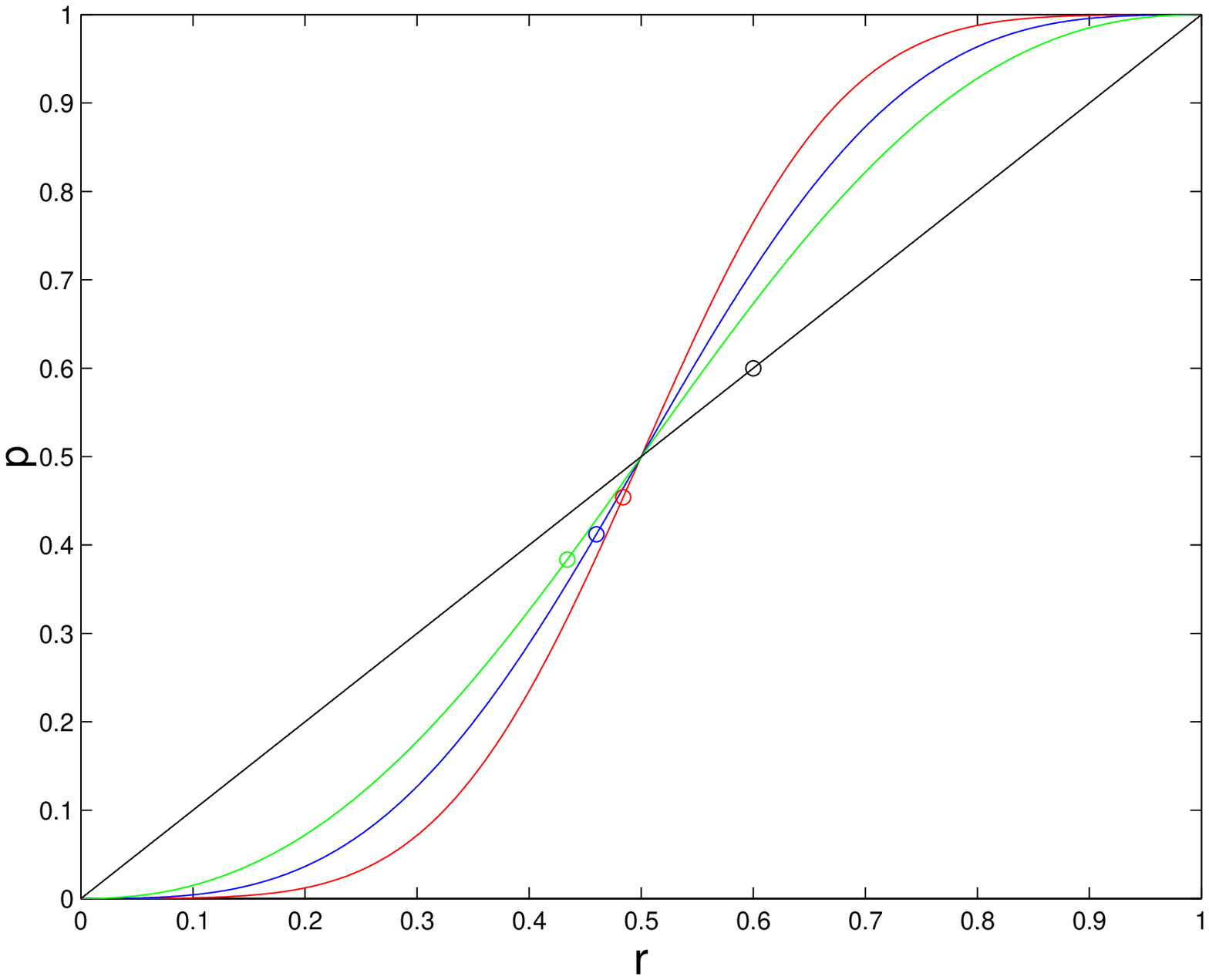}
\end{center}
\caption{\small {The concentration $p$ of filled sites as a
function of cutoff $r$ for random model (black curve), and three
different self-affine models: $H=0.2$ (green), $H=0.5$ (blue) and
$H=0.8$ (red). In each case, the value of $r^{eith}_c$  at
$p^{eith}_c$ has been marked (circles).}} \label{fig.2}
\end {figure}
\begin{figure}
%\bigskip
%\bigskip
\begin{center}
\includegraphics[width= 10 truecm]{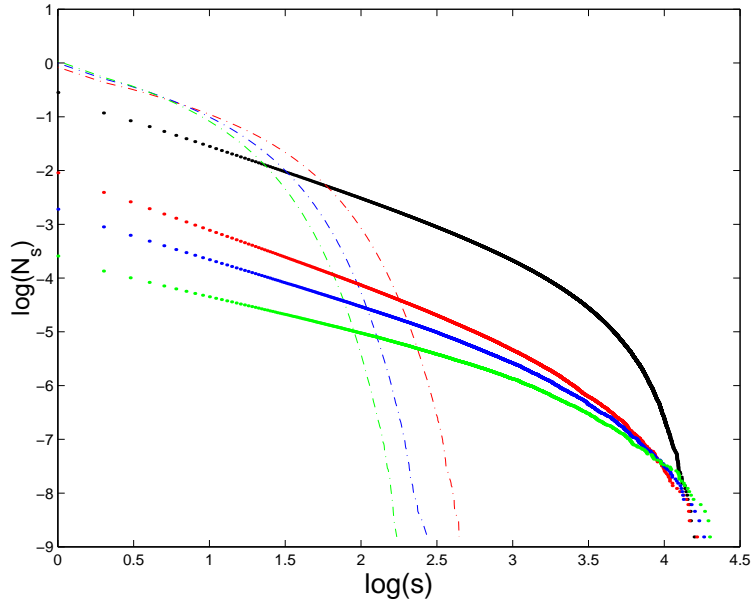}
\end{center}
\caption{{\small Variation of $N_s= \sum_{k \geq s} n_{k}$, where
$n_k$ is the normalized cluster number of size $k$ for
percolation in random model (black) and percolation on either
direction in three different self-affine models: $H=0.2$ (green),
$H=0.5$ (blue) and $H=0.8$ (red). The dash-dot curves are the
$N_s$'s in random model when the concentration of filled sites is
equal to the $p_c^{eith}$ of self affine model when $H=0.2$
(green), $H=0.5$ (blue) and $H=0.8$ (red). In all cases we have
only counted the internal percolation
clusters of $256 \times 256 $ square lattices.}} %\label{fig.}
\end {figure}
To produce the percolation clusters of the self-affine model we
have used the percolation thresholds provided in Ref.~[16]. In
Fig.2 we have shown our estimations of the concentrations $p$ of
filled sites as a function of cutoff $r$ for three different
values of Hurts exponent, $H=0.2$, $H=0.5$ and $H=0.8$. For
random model, $p=r$, but as it is seen this is not the case for
self-affine models. In fact, the fraction of occupied sites at
both $p_c^{eith}$, and $p_c^{spec}$ in self-affine models is less
than the corresponding values of random model. This is because
compared to the random case, the existence of self-affine
long-range correlations is in favor of formation of larger
clusters. Calculation of $N_s= \sum_{k \geq s} n_{k}$, where
$n_k$ is the total number of clusters with size $k$ per lattice
sites (fig.3) for the case of percolation on either direction
shows this phenomenon clearly. For comparison, the $N_s$'s in
random model when the concentration of filled sites is equal to
$p_c^{eith}$ of self affine models with $H=0.2$, $H=0.5$ and
$H=0.8$ are shown too.
%\bigskip

Our numerical results for the values of the mean anisotropy
parameters at percolation transition in either direction (rule $
\mathcal R_0$) and percolation transition in specified direction
(rule $\mathcal R_1$)  have been appeared in fig.4 and fig.5
respectively. For comparison the corresponding values for random
percolation have been also included. Interestingly, the
differences between these two groups of curves are almost
negligible, although the difference between $p_c^{eith}$ and
$p_c^{spec}$ is significant. In fact more investigations suggest
that unlike random model, the shape of clusters of the self-affine
model do not vary significantly with concentration $p$ of the
filled sites. We may conclude that the introduction of  isotropic
self-affinity subsidizes (and even maybe removes) the dependence
of the number of a specific cluster configuration to
concentration of filled sites.
\begin{figure}
\bigskip
\begin{center}
\includegraphics[width= 10 truecm]{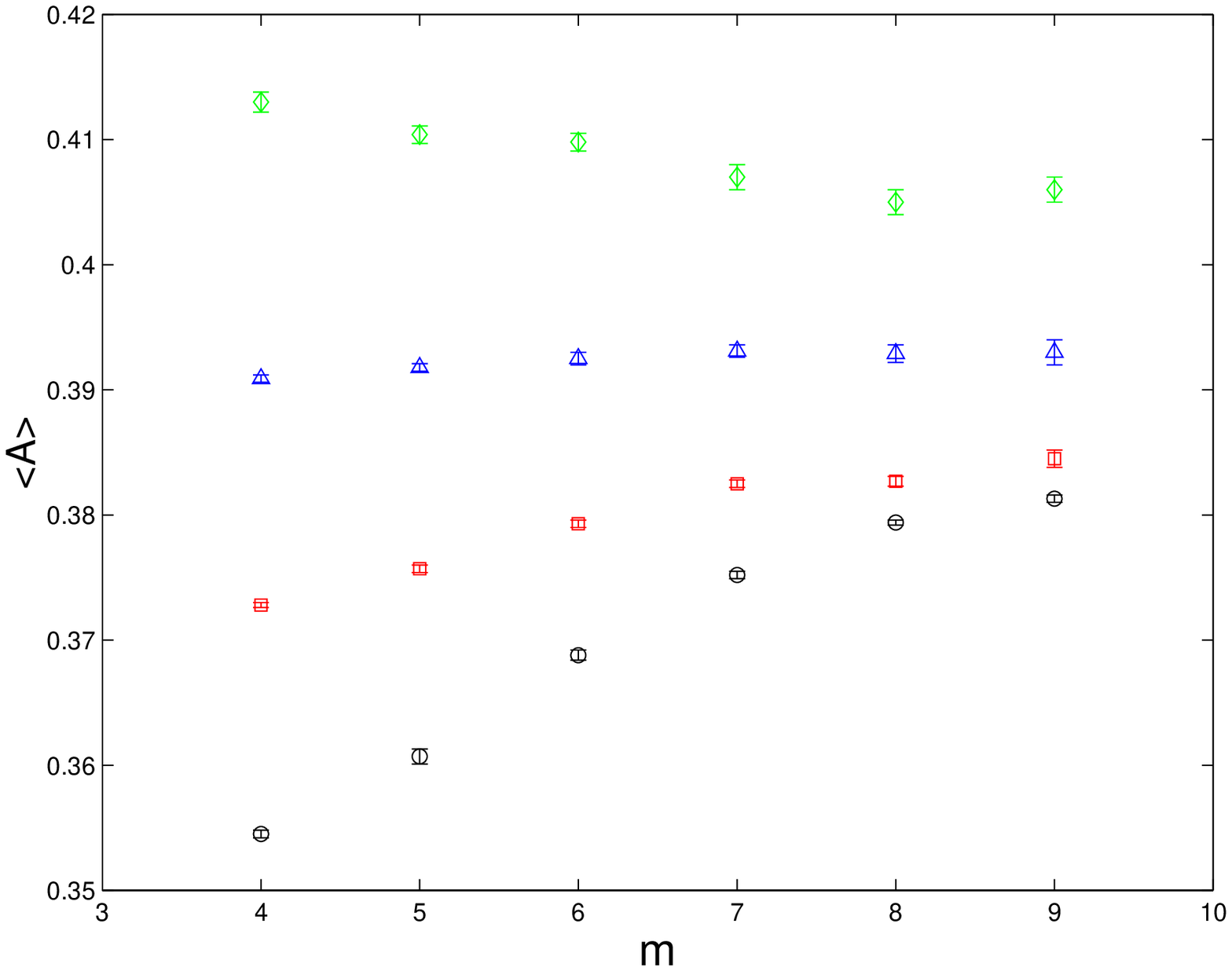}
\end{center}
\caption{\small{The estimated values of mean anisotropy parameters
at $p_c^{eith}$ for self-affine models with $H=0.2$ (green),
$H=0.5$ (blue) and $H=0.8$ (red). The results have been lumped
together at the block centers of size $[2^{m-1},2^m]$. For
comparison the corresponding values for random percolation at the
percolation threshold have been also included (black).}}
\end {figure}
\begin{figure}
%\bigskip
\begin{center}
\includegraphics[width= 10 truecm]{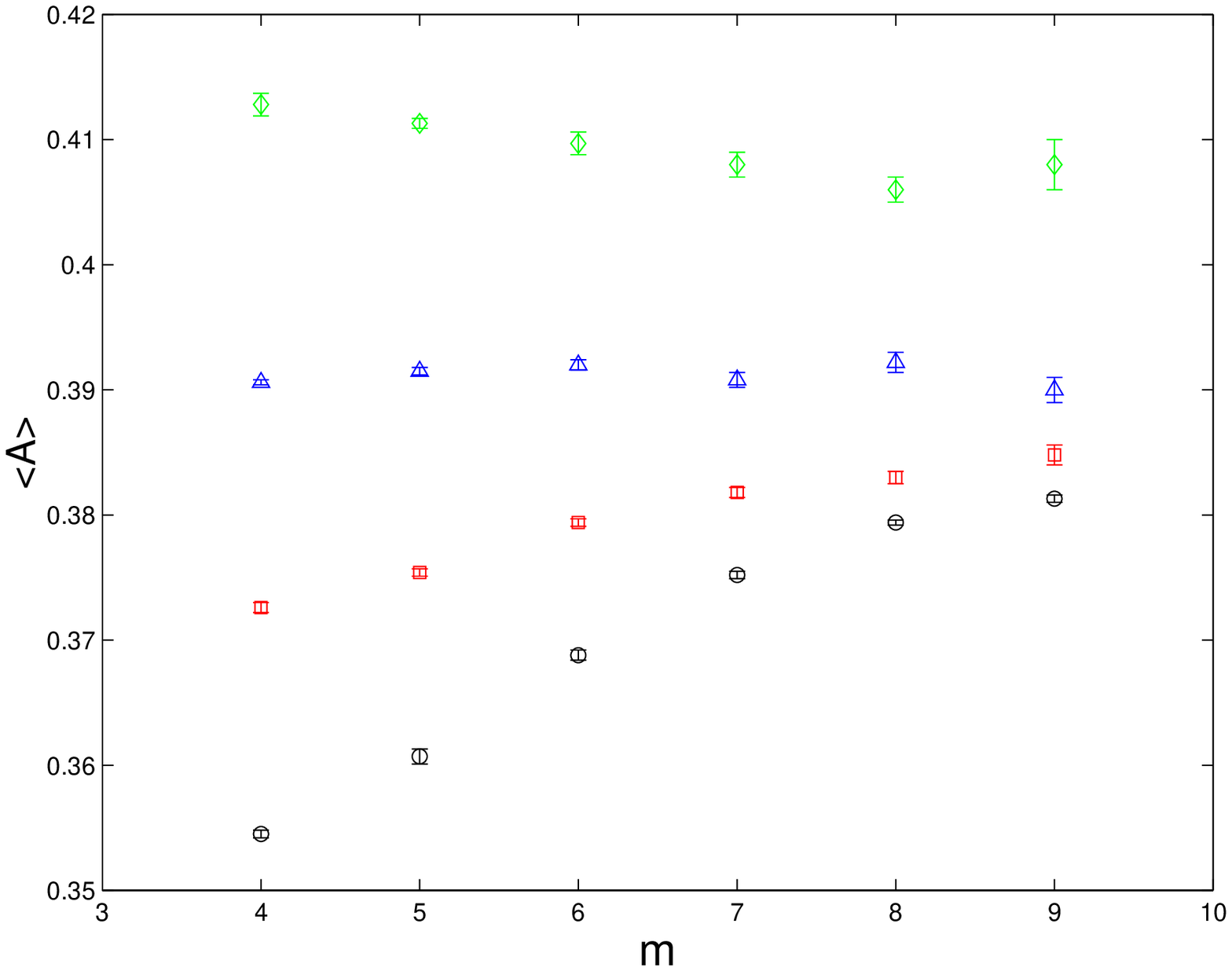}
\end{center}
\caption {\small{The estimated values of mean anisotropy
parameters at $p_c^{spec}$ for self affine models with $H=0.2$
(green), $H=0.5$ (blue) and $H=0.8$ (red). The results have been
lumped together at the block centers of size $[2^{m-1},2^m]$. For
comparison the corresponding values for random percolation at the
percolation threshold have been also included (black).}}
\end {figure}

One can observe that another important effect of self-affinity is
to increase the isotropy of percolation clusters. Meanwhile, the
difference between the anisotropy parameters of different $H$'s
and random percolation clusters are not significant especially
when $H\rightarrow 0$. This result is in agreement with the
previous findings which say that in two dimensions, the scaling
properties of self-affine model does not differ so much from
random model and the difference between the critical exponents of
two models vanishes as $H\rightarrow 0$~\cite{sam}.

While the presence of both kind of long-range correlations
increases the isotropy of self affine percolation clusters, there
is still a significant qualitative difference between the effect
of persistent and anti-persistent correlations on the shape of
clusters. Considering the relation:
\begin{equation} \langle A_s \rangle\cong A_{\infty}
[1+(a_{min}-a_{max})s^{-\theta}+(b_{\min}-b_{max})s^{-1}]
\end{equation}
we can deduce from the behaviour of anisotropy parameters that
the sign of correction to scaling term,
$(a_{min}-a_{max})s^{-\theta}$, depends on $H$: it is positive for
$H<0.5$ and negative for $H>0.5$. At $H=0.5$ this term is almost
zero. Hence, for anti-persisting long-range correlations the
correction to scaling coefficient $a_{min}$ is larger than
$a_{max}$, which means in this case $R^2_{min}$ grows faster than
$R^2_{max}$. Again, we see that anti-persistent self-affine models
shows more similarity with random model (see fig.4 or fig.5). In
the presence of persisting correlations $R^2_{max}$ grows faster
than $R^2_{min}$ and the anisotropy parameter decreases with $s$.
Naturally, for the case $H=0$ which separates these two kinds of
correlations, the growth rate of $R^2_{min}$ and $R^2_{max}$
should be equal.

\section{ Concluding remarks}

 We calculated numerically the anisotropy in the shape of equilibrium
 percolation  clusters in long range self-affine correlated
 square lattices and  observed some interesting results.  We saw that
 compared to random percolation,
  the shape  of percolation clusters become slightly more
  symmetric and the exact value of anisotropy parameter is a
  function of Hurts exponent. On the other hand,
 the type of  correlations in the medium property, determines the sign of
 non-analytical  correction to scaling contribution into the anisotropy
 parameters of clusters. In accordance with the
 previous studies we also observed that  the percolation clusters of
 FBM  models become much more  similar to those of random models as $H\rightarrow 0$.

Finally, we mention again that, the size of largest cluster that
we may  consider in our analysis is restricted by the lattice
size.  To estimate the accurate values of $A_\infty$ or
non-analytical correction to scaling exponents one should extend
the simulation method to larger values of $L$.

\noindent The author would like to thank M. Sahimi for useful
hints.

% \section*{References}

\end{document}